\documentclass[10pt]{article}
\usepackage[a4paper, total={6in, 8in}]{geometry}
\title{TSDF: A simple yet comprehensive, unified data storage and exchange format standard for digital biosensor data in health applications}
\usepackage{graphicx,color,soul,natbib,rotating,moreverb,amssymb}
\usepackage[hidelinks]{hyperref}
\usepackage{comment}
\usepackage{import} %for the pdf_tex
\usepackage[table,xcdraw]{xcolor}

\author{
	\parbox{\linewidth}{\centering
	Kasper Claes$^1$, Valentina Ticcinelli$^1$, Reham Badawy$^2$, Yordan P. Raykov$^3$, Luc J. W. Evers$^4$, Max A. Little$^2$\\
	\small $^1$ UCB Pharma, 1070 Brussels, Belgium; kasper.claes@ucb.com, valentina.ticcinelli@ucb.com\\
	\small $^2$ School of Computer Science, University of Birmingham, Birmingham, UK; rehambadawy@hotmail.com, maxl@mit.edu\\
	\small $^3$ University of Nottingham, UK; yordan.raykov@nottingham.ac.uk\\
	\small $^4$ Department of Neurology, Donders Institute for Brain, Cognition and Behavior, Radboud University Medical Center, Nijmegen, The Netherlands; luc.evers@radboudumc.nl
}
}
\begin{document}

\maketitle
\begin{abstract}
Digital sensors are increasingly being used to monitor the change
over time of physiological processes in biological health and disease, often using wearable devices. This generates very large amounts of digital sensor data, for which a consensus on a common storage, exchange and archival data format standard, has yet to be reached. We pose a series of format design criteria and review in detail existing storage and exchange formats. When judged against these criteria, we find these existing formats lacking, and propose Time Series Data Format (TSDF), a unified, standardized format for storing all types of physiological sensor data, across diverse disease areas. TSDF is simple and intuitive, suited to both numerical sensor data and metadata, based on raw binary data and \texttt{JSON}-format text files, for sensor measurements/timestamps and metadata, respectively. By focusing on the common characteristics of diverse biosensor data, we define a set of necessary and sufficient metadata fields for storing, processing, exchanging, archiving and reliably interpreting, multi-channel biological time series data. Our aim is for this standardized format to increase the interpretability and exchangeability of data, thereby contributing to scientific reproducibility in studies where digital biosensor data forms a key evidence base.
\end{abstract}
\def \figwidth {\columnwidth}
\def \figwidthSmall {1.0\columnwidth}
\def \figwidthLarge {\columnwidth}

\section{Introduction}
\begin{figure*}
\centering
  \includegraphics[width=\textwidth]{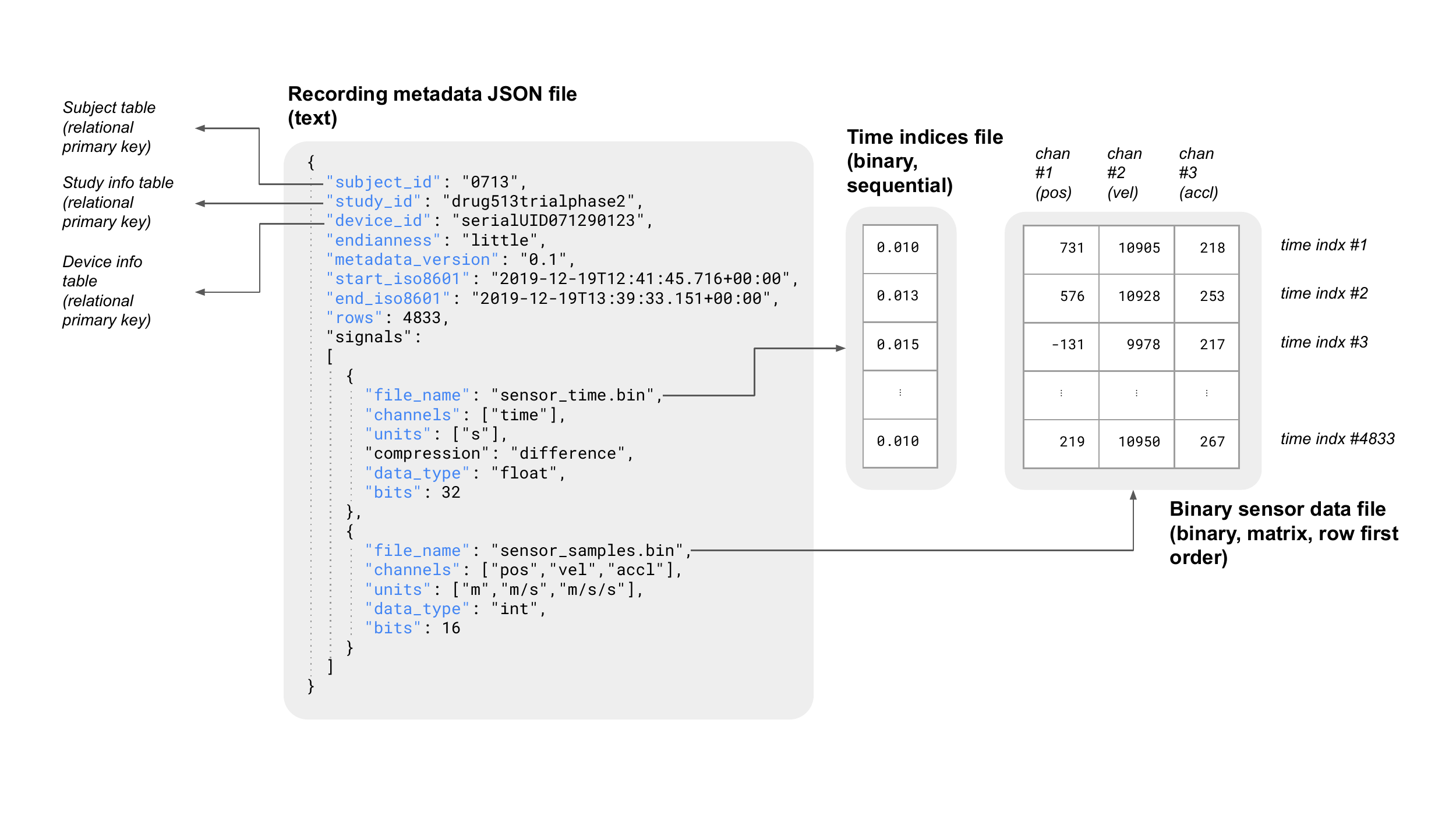}
  \caption{Typical example of our proposed
sensor data standard, TSDF. Sensor data recording from a single participant in a clinical drug trial, described with minimal mandatory metadata (fields in blue), stored
in three separate files: the metadata file, a sensor time index binary
file (sensor$\_$time.bin), and the sensor data samples binary file
(sensor$\_$samples.bin). The metadata file is a \texttt{JSON} text file \citep{bray2014javascript}, readable
and editable in any text editor. The binary files are stored as binary numbers in sequential vectors (e.g. matrices) in row first order. The subject$\_$id, study$\_$id and device$\_$id fields of the metadata are primary keys in associated auxiliary relational database
tables for further subject, study and device details (specification
for these auxiliary tables is not part of the proposed data format
standard).}
  \label{recordingMetadata}
\end{figure*}

Digital biosensor data can contribute to the prevention and diagnosis of disease, and may provide actionable treatment insights for both safety and efficacy \citep{badawy2019metadata}. Across diverse disease areas, monitoring at a high sampling rate (at home or in clinic, longitudinally or for short periods) provides insights into the patients' condition. Such recordings generate large amounts of data. The universal adoption of an appropriate, standardized data storage and exchange format across application areas, is desirable for multiple reasons; but primarily because it would greatly facilitate data sharing and re-use.

Unfortunately the digital health research community has yet to reach a consensus on a unified, open format for storing and exchanging biosensor data. For scientific reproducibility and exchange, it is desirable that this format becomes an open standard. This would enable interoperability and long term archival preservation, saving substantial time, cost and wasted effort compared to the current situation where a variety of different, proprietary and/or mutually incompatible data formats are used. An open standard requires public specification. Our aim in this paper is to present such a standardized format, and to do so we reuse existing open standards both for metadata and numerical data files, in order to ensure readability (now and later), software platform independence and vendor neutrality \citep{piper18}. We also aim to ensure that the format is simple and entirely intuitive, so that complex software tools and documentation are not needed.

Getting to a workable consensus on an open data standard is a long-term process, requiring compromise between all stakeholders. A good candidate for this conversation is a format that is both generalisable to the needs of different users and flexible enough to apply in every application. A standard needs to be both flexible and simple to use. We start from the observation that all physiological time series are similar, whether originating from e.g. electrophysiology, inertial devices, electrodermal activity or photoplethysmography: all consist of a series of sensor measurements with a relatively high, and often irregular (non-uniform), sampling rate. Furthermore, a data format  should not be specific to a particular (animal) species, nor to a specific disease area. Hence our proposed format is designed to fit all sensor types, species and health contexts.

Our design choices are optimised for raw sensor data, typically captured at relatively high and/or irregular sampling rates, such as electrocardiogram (ECG) data. However a useful data format in this context should be able to represent both raw and \textit{processed} data (also called \textit{derived} data, typically at a much lower sampling rate). An obvious example would be heart rate derived from ECG. Due to the simplicity of the proposed format, derived feature data can also be stored in the same way.

For the sake of maintaining simplicity, we consider certain kinds of data which, while being time-ordered, require very different technical specifications to sensor time series, such as surveys, demographic data, trial protocol sequential information etc. Nevertheless, these other types of data are critically important to maintaining the proper \textit{interpretation} of the sensor time series data captured during experimental studies. For this reason, our proposal splits the relevant metadata for a study, into \textit{sensor-specific} and other, metadata, and provides minimal \textit{relational keys} in the sensor-specific metadata which link to database tables storing this wider, contextual data. This approach keeps the sensor metadata concise which is critical for data access for algorithmic processing and storage purposes, yet retaining the ability to reconstruct the context of the digital sensor data.

The structure of the paper is as follows. Section \ref{design} elaborates on the design criteria for metadata and numerical data formats; Section \ref{related} then provides a comprehensive overview of previous, related work; finally, Section \ref{proposal} details TSDF, the proposed format.

\section{Design criteria} \label{design}
The discussion above leads to specific design criteria, which distinguish this particular storage format from other kinds of problems, and are discussed in detail below. Along with the technical requirements emerging from the need to process such large scale sensor datasets algorithmically in an efficient way, a reproducible scientific data format must be \textbf{sustainable} as formalized in the FAIR principles: \textit{findable}, \textit{accessible}, \textit{interoperable} and \textit{reusable} \citep{wilkinson2016}.

\begin{itemize}
\item \textbf{Storage efficiency}:  sampling is often non-uniform, and in this case there is the need to associate a time index with every sample. Sensors typically generate many values per second of both measurements and timestamps. A storage efficient format for both is required to make optimal use of device battery life, memory, and transfer bandwidth.

\item \textbf{Comprehensiveness}: complete metadata is needed in order to ensure correct interpretation of the context of the measurements, for example subject, device and study-related information, and at the other extreme, measurement-related details e.g. sampling frequency, numerical sensor channel scale factors, measurement bit depth and binary representation.

\item \textbf{Flexibility}: while a minimal subset of the metadata needs to be mandatory, there needs to be flexibility, for example, in allowing special fields or annotations for specialized applications. Flexibility is also relevant at the level of the entire data set: while on a server different parts of a data set can be kept as separate files in the same folder, it is advisable to enclose them in a single container file for data transfers.

\item \textbf{Simplicity}: both conceptual and technical simplicity are required. A conceptually simple data structure design has many advantages, for example, should the integrity of the data become degraded or the data format documentation have become lost, the whole can be reconstructed easily with minimal, if any, additional information. Technical simplicity minimizes the effort needed for building, maintaining and reconstructing software tools for archival purposes. Self-explanatory names for the metadata fields facilitate data sharing and reuse.

\item \textbf{Accessibility}: since biosensor data is often large-scale, the format should support native random access for efficient handling of a specific section of a file without loading it in its entirety. Loading the data into memory should be as fast as possible, avoiding unnecessary parsing and/or decompression. Appropriate file sizes are also crucial: many small files complicate filing system management, while very large files typically contain only a small fraction of relevant data \citep{sauermann17} and increase the potential for catastrophic loss if a single file is damaged \citep{stead16}. 
\end{itemize}

These are ideals and cannot all be fully satisfied simultaneously. For instance, storage efficiency is in competition with accessibility, since sophisticated compression would optimize storage but require decompression on loading which would be computationally complex and therefore introduce an overhead on loading. Thus, the design of any such format involves trade-offs.

\subsection{Numerical data}
Considerations about numerical data storage apply to both sensor measurements and timestamps. In some circumstances, it can be convenient to keep the timestamp and sensor measurements in the same file. However, particularly when the numerical time format is essentially incompatible with the sensor measurement format, or with increasing dataset length, it is usually better to split the time information from the measurements, as this often allows the  timestamps to have a simple, yet highly compact representation.

\subsubsection{Timestamp representation}
As sampling of biosensor data is typically non-uniform, a timestamp usually needs to be associated with every measurement instant. As this leads to a large quantity of data, not only the sensor measurement values but also timestamps associated with them, need an efficient numerical storage format.

Unsigned integers (\texttt{uint}) are usually the most suitable format for this purpose, as the time-axes are typically built by sampling at multiples of a base unit from the start of the recording. For example, assuming an average sampling rate of up to 1kHz, millisecond resolution is often sufficient. In this case a 32 bit \texttt{uint} is able to encode a recording of up to 50 days' duration.

According to considerations such as recording session duration, random access granularity and desired precision of the timestamps, several common time encoding methods are appropriate:

\begin{itemize}
        \item \textit{relative}: time elapsed since the start time of the recording. This allows for random access and is a space efficient encoding, as it starts from zero in each file.
        \item \textit{absolute}: for example, Unix time with millisecond precision. This enables very simple random access but is not as storage efficient as relative timestamps.
        \item \textit{difference}: time elapsed since the previous sample. This is useful for very long recordings with large time differences between samples. However, it makes random access more complex because the current time needs to be reconstructed from previous timestamps.
        \item \textit{uniform}: when the sampling is uniform, there is no need for measurement-by-measurement timestamps. The only information required is the start time of the recording, and the sampling frequency, in order to determine the timestamp for every measurement instant.
\end{itemize}

\subsubsection{Measurement representation}
To ensure storage efficiency and flexibility, the measurement representation needs to be compact and configurable. The parameters required to parse it need to be specified in the metadata, alongside the corresponding units \citep{sauermann17}. Format flexibility is required because, for both measurements and timestamps, the optimal storage format depends on the specifics of the data. For reasons of interoperability it is wise to limit the choice to internationally-standardized number formats, for example the IEEE 754 floating-point format. In addition, a \textit{scale factor} field in the metadata allows further storage efficiency by adapting the number format to best fit the resolution and range requirements \citep{sauermann17}.

\subsection{Metadata}
Each field in the metadata structure is either \textit{mandatory, desirable} or \textit{additional} \citep{garcia14}. Mandatory fields are required to fully specify the data: these are kept to a minimum. Desirable fields capture commonly encountered information, this is the bulk of the proposed metadata structure, and encompasses a very wide variety of use cases. Additional fields are suggestions for variables which may address other, as yet unforseen applications.

\subsubsection{Versioning}
A mandatory metadata version field accommodates unanticipated changes to the metadata, which may be required in future, e.g. in response to changes in technology.

\subsubsection{Readability}
When defining a metadata structure, it is important to consider that, ideally, every file should be both human- and machine- readable. To satisfy the former, a text format is a logical choice. For the latter, free text is best kept to a minimum in favour of structured fields that indicate a one-to-one relationship between each text string and the associated concept. Where free text is unavoidable, one might, for example, use a descriptive comment together with a \textit{uniform resource identifier} (URI) to an \textit{ontology} that unambiguously defines the concept.

\subsubsection{Automated querying}
One of the main advantages of machine-readable metadata is that a database can be automatically constructed from it. A script reads the \texttt{JSON} files and parses the metadata into a \textit{relational database}. This is a well-defined technology for handling manipulations of this kind of, essentially tabular, data, stored in multiple tables which are linked together using so-called \textit{keys}. This makes the metadata searchable using \textit{structured query language} (SQL) to answer sophisticated questions about the data, for example, cross-sectioned by subject, device, sensor type, date or time of the day, in complex combinations. Ideally, in order for all entries related to a concept to be robustly accessible despite possible multiple naming choices, fields that relate to notions such as units, sensor types and number formats can be unambiguously defined using repositories that combine various controlled vocabularies, such as \href{https://bioportal.bioontology.org}{Bioportal} and the \textit{Unified Medical Language System}. These link among others to other, subject-specific ontologies such as Snomed, MeSH, ICD-10, CDISC and HL7.

\subsubsection{Time representation}
Time in the metadata schema is expressed as a structured string, as standardized by ISO 8601, making it both machine- and human- readable. This includes arbitrary precision for fractions of a seconds: its format is \texttt{yyyy-mm-ddThh:mm:ss.SSS±hh:mm}. This string clarifies both the absolute time (for chronological ordering), and the local time, which is physiologically relevant as it relates to the circadian rhythms and other behaviours of the subject. The standard also includes an encoding for situations when either of these is unknown. If, for instance, one only knows the \textit{coordinated universal time} (UTC) without the time zone offset, a \texttt{Z} is used in the place of the \texttt{±hh:mm} string; if only the local time is known, the time zone information \texttt{±hh:mm} is simply omitted.

\begin{sidewaystable*} % <-- HERE
\centering
\scriptsize
\begin{tabular}{p{.06\linewidth} p{.06\linewidth} p{.06\linewidth} p{.06\linewidth} p{.06\linewidth} p{.06\linewidth} p{.06\linewidth} p{.06\linewidth} p{.06\linewidth} p{.06\linewidth} p{.06\linewidth} p{.06\linewidth} p{.06\linewidth} p{.06\linewidth}}\hline
Format & Parse-free random access & Non-uniform timestamps & Text & Binary & Flexible metadata & Hierarchical metadata & Self-explanatory & Compact numerical representation & Flexible numerical representation & Parse-free data loading & Formal specification & Open standard numerical representation & Machine readable specification \\\hline
CSV &               x &          \checkmark & \checkmark & x &          x &          x &          \checkmark & x &          \checkmark & x &          x &          \checkmark & x\\
\href{https://www.json.org/json-en.html}{JSON} &              x &          \checkmark & \checkmark & x &          \checkmark & \checkmark & \checkmark & x &          \checkmark & x &          \checkmark & \checkmark & \checkmark\\
\href{https://www.w3.org/TR/xml}{XML} &               x &          \checkmark & \checkmark & x &          \checkmark & \checkmark & \checkmark & x &          \checkmark & x &          \checkmark & \checkmark & \checkmark\\
\href{https://www.rfc-editor.org/rfc/rfc8949.html}{CBOR} &              x &          \checkmark & x &          \checkmark & \checkmark & \checkmark & x &          \checkmark & \checkmark & x &          \checkmark & \checkmark & x\\
\href{https://avro.apache.org/docs/1.10.2/spec.html}{Avro} &              x &          \checkmark & x &          \checkmark & \checkmark & \checkmark & x &          \checkmark & \checkmark & x &          \checkmark & \checkmark & x\\
\href{https://docs.hdfgroup.org/hdf5/develop/_f_m_t3.html}{HDF5} &              x &          \checkmark & x &          \checkmark & \checkmark & \checkmark & x &          \checkmark & \checkmark & x &          \checkmark & \checkmark & x\\
\href{https://www.loc.gov/preservation/digital/formats/fdd/fdd000001.shtml}{WAV} &               \checkmark & x &          x &          \checkmark & \checkmark & x &          x &          \checkmark & x &          \checkmark & x &          x & x\\
%\href{https://www.loc.gov/preservation/digital/formats/fdd/fdd000012.shtml}{MPEG-3} &            x &          x &          x &          \checkmark & \checkmark & x &          x &          \checkmark & x &          x &          \checkmark & x \\
\href{https://www.edfplus.info/}{EDF/EDF+} &          \checkmark & x &          \checkmark & \checkmark & x &          x &          x &          \checkmark & x &          \checkmark & x &          x & x\\
\href{https://arxiv.org/abs/cs/0608052}{GDF} &               \checkmark & x &          \checkmark & \checkmark & x &          x &          x &          \checkmark & x &          \checkmark & x &          \checkmark & x\\
\href{https://main.ieeg.org/sites/default/files/MEF_Format.pdf}{MEF} &               x &          \checkmark & x &          \checkmark & x &          \checkmark & x &          \checkmark & x &          x &          x &          x & x\\
\href{https://developers.google.com/protocol-buffers/}{Protocol Buffers} &  x &          \checkmark & x &          \checkmark & \checkmark & \checkmark & x &          \checkmark & \checkmark & x &          \checkmark & \checkmark & x\\
\href{https://github.com/google/flatbuffers}{Flatbuffers} &       x &          \checkmark & x &          \checkmark & \checkmark & \checkmark & x &          \checkmark & \checkmark & x &          \checkmark & \checkmark & x\\
\href{https://meg.univ-amu.fr/wiki/AnyWave:ADES}{ADES} &              \checkmark & x &          \checkmark & \checkmark & x &          x &          \checkmark & \checkmark & x &          \checkmark & x &          x & x\\
\href{https://docs.movisens.com/Unisens/UnisensFileFormat/#overview}{Unisens} &           \checkmark & x &          \checkmark & \checkmark & x &          x &          x &          \checkmark & \checkmark & \checkmark & x &          x & x\\
TSDF (proposal) &  \checkmark & \checkmark & \checkmark & \checkmark & \checkmark & \checkmark & \checkmark & \checkmark & \checkmark & \checkmark & \checkmark & \checkmark & \checkmark\\
\hline
\end{tabular}
\caption{Formats labelled as both text and binary have mixed text metadata with binary numerical data storage.\\
For ``parse-free" formats, the original raw numerical data may be reconstructed directly from storage, with no additional computational steps (i.e. through Direct Memory Access transfers).\\
``Flexible metadata" formats allow unrestricted metadata fields.\\
``Self-explanatory" formats require no additional documentation or software tools in order for the data to be reconstructed and used.}
\end{sidewaystable*}

\section{Related data format standards} \label{related}
There are a number of technical possibilities for storing numerical data and metadata, some of which have already been implemented as data formats. Below, we evaluate the suitability of these possibilities in terms of the design criteria described in the previous section.

\begin{itemize}
\item \textbf{Text only}:  both numerical data and metadata can be stored in a text format, such as \textit{comma separated values} (CSV), \textit{extensible markup language} (XML) or \textit{JavaScript object notation} (\texttt{JSON}).

\item \textbf{Binary only}:
  \begin{itemize}
  \item \textbf{Specialized medical data format}: both metadata and data are encoded in a single, dedicated file (e.g. EDF+, GDF and MEF). It can
be sensor-type dependent or independent.

  \item \textbf{Serialization container}: a data structure, stored in computer memory, is \textit{streamed} to a \textit{container} file, with a specification of how to encode the data in this container (such as the Hierarchical Data Format).
  \end{itemize}

\item \textbf{Mixed text and binary}:
  \begin{itemize}
  \item \textbf{Hybrid file}: a single file where a \textit{header} contains the metadata in text format, with the remainder of the file, in binary format, containing the numerical data. The major difficulty of this arrangement is it complicates input/output operations: one needs to parse headers of variable length in order to access the numerical data, severely compromising random access performance \citep{greenfield15}. Furthermore, without access to the header format, it is often extremely difficult to extract the numerical data alone.

  \item \textbf{Separate files for metadata and numerical data}: e.g. metadata in text, numerical data in raw binary files. This solution ensures metadata readability while being storage-efficient for the numerical data. The only major risk of this approach is of the numerical data and metadata files becoming separated through inadvertent mishandling of a dataset.
  \end{itemize}
\end{itemize}

In addition to this technical subdivision, the medical scope of a data format can differ. Concrete examples of the different approaches are discussed in this section.

\subsection{Text only}
A text format has the inherent advantage of being readable by humans without the need for any software besides a ubiquitous text editor. However, any text format is extremely inefficient for storing high sampling rate data, since the resulting file size would be at least a multiple of the corresponding binary file, with large amounts of redundancy. This excess causes energy and resource inefficiencies, such as lengthening data transfer times and placing unnecessary burdens on both device and machine memory \citep{piper18}. Standard dictionary-based compression schemes (such as Lempel-Ziv) only partially solve this data size issue, as it makes random access to the data impractical. For these reasons, a text format can be an adequate choice for saving numerical data in case of very low (e.g. less than 1Hz) sampling rates and then only for very small datasets.

From experience, use of the CSV format is widespread for numerical data. This is probably due to its compatibility with spreadsheet software programs, which arrange the data in tables and provide simple calculations and visualisation tools. However, this format has a number of serious flaws. First, there is no single agreed-upon standard: the format depends on local parameters which leads to different interpretations of the full stop, comma, semicolon and string terminator characters, which in turn leads to the high risk of mismatching rows and/or columns. In addition, as with many informally-specified text formats, the field length is variable, hence random access is not possible and computationally expensive parsing is required. Also, no metadata hierarchy is possible in CSV format, and it is not clear which convention applies when the string terminator must appear as part of a string.

\texttt{JSON} is a lightweight, data-interchange text format, originating as a domain specific language for describing data structures. It has the advantage of being human-readable, but it is also structured and therefore easily parsed automatically. \texttt{JSON} is built on two universal structures supported by most modern computer languages, i.e. a collection of name/value pairs (i.e. a \textit{dictionary}) or an ordered list of values (i.e. an \textit{array}). The non profit organization Open mHealth stores both numerical data and metadata in \texttt{JSON} format. This format is particularly appropriate for storing structured information, such as metadata information. However, as with any other text format, it impedes random access and is inefficient for numerical data, requiring at least four times the data volume compared to binary storage.

\subsection{Sensor-type specific binary/mixed}

In cardiological applications, a number of formats have been proposed specifically for electrocardiogram (ECG) data, for example \href{https://www.en-standard.eu/csn-en-1064-health-informatics-standard-communication-protocol-computer-assisted-electrocardiography/}{SCP-ECG} (aka OpenECG), \href{https://dicom.nema.org/dicom/supps/sup30_lb.pdf}{DICOM-ECG} (aka Waveform) and HL7 \href{https://www.hl7.org/implement/standards/product_brief.cfm?product_id=70}{aECG}. None of these three formats are openly available at no cost. In addition these are specifically designed for ECG data, the specifications for these formats tend to be overly complicated, and software support is currently rather limited \citep{bond11,badilini18,stead16}.

In neuroscience/neurology applications, a number of initiatives exist exclusively for electrophysiology data. Some formats are based on an existing binary container e.g. HDF5 (Hierarchical Data Format, version 5). Two examples are: Neurodata Without Borders, with a focus on cellular neurophysiology data \citep{teeters15,ruebel19}, and Brainformat, for electrophysiology data \citep{ruebel16}.

Other open data formats specifically designed for electrophysiology use a specialized binary structure, which we describe next. Multiscale Electrophysiology Format (MEF) allows for hierarchically-structured  numerical data and metadata, and supports encryption and compression \citep{brinkmann09} as part of the standard. It uses a fixed-length metadata header with predefined fields, and therefore lacks metadata flexibility for potential use with sensor data other than electrophysiology. It allows for inclusion of non-sensor data types such as video and events, which are frequently paired to electrophysiological measurements in experiments. However, such files have other technical requirements and would be better  encoded in a data format more suited to their specific characteristics. Furthermore, a compression/decompression format such as MEF needs detailed documentation due to its relative complexity \citep{stead16}.

European Data Format (EDF or EDF+) is a widely-used format for storing electroencephalogram (EEG) data which uses a specialized binary encoding. It does not allow for compression or encryption \citep{stead16}, and is relatively hard to implement outside of the specific software packages that are designed to handle it \citep{mihajlovic15}. It does not follow international open standards: for example, the date format is not compliant with ISO 8601. EDF+ does not allow storing a timestamp associated with each measurement, and therefore assumes uniform sampling \citep{kemp03}. Due to its container-style structure, adding or removing information involves recreating the entire file in EDF+ \citep{lambert15}. Furthermore, the numerical data can only be stored in a single, fixed format, i.e. 16 bit integers, hence severely limiting the trade-off between range and resolution  \citep{beier19,stead16}.

General Data Format (GDF) emerged as an effort to overcome the shortcomings of EDF+ \citep{Schloegl09}. It includes metadata outside of the sensor domain such as events, demographics, manual quality control annotations, and clinical data. However, this metadata is highly specialized and omits information that is essential for defining the context of the measurements. Furthermore, the very specialized and therefore largely irrelevant metadata, that are included in this format, are all mandatory, because they are stored in specific bits in a fixed length binary header \citep{sauermann17,schloegl13}.

\subsection{Sensor-type independent binary/mixed}

\subsubsection{Specialized binary}
Medical Waveform Encoding Rules (\href{http://www.mfer.org/en/index.htm}{MFER}) is a sensor-type independent format that was created with the same purpose as this publication: to serve for all physiological time series across health indications and species. It has an associated international standard (ISO 22077-1:2015) and has a small file size. Unfortunately, the format has not been widely adopted, presumably because such adoption requires familiarity with its copious and complex documentation, as it makes use of complex data structures such as sequences, blocks and frames. It does not use existing standards, e.g. units are redefined. It also does not allow associating a timestamp with each measurement.

\subsubsection{Text metadata and binary numerical}
Another example of a sensor-type independent format for time series data is AnyWave DEScriptive format (\href{https://meg.univ-amu.fr/wiki/AnyWave:ADES}{ADES}), which consists of a text file containing the metadata, and a binary dump file with numerical data. The data is channel-multiplexed, i.e. arranged chronologically, hence enabling random access. Movisens implemented a similar format called \href{https://docs.movisens.com/Unisens/UnisensFileFormat/}{Unisens}, with XML metadata and binaries for the measurements. Unfortunately these formats do not allow associating a timestamp with each measurement. Another drawback is that the metadata does not include fields critically important for context information such as a subject identifier, study identifier, device identifier, or recording start timestamp, among other omissions.

\subsubsection{Serialization containers}
There are many serialization formats that are widely used for other scientific applications, which may be suitable for physiological time series. One of the main drawbacks of such formats relates to long-term archival data preservation. These containers have complicated and highly flexible encoding schemes with the risk that when technology advances, the software tools and the platform supporting those tools becomes obsolete, so the format falls into obscurity with the potential that the infrastructure needed to decode these formats can be lost.

Concise Binary Object Representation (CBOR) is essentially a binary version of \texttt{JSON}: it is geared towards name-value pairs, but it also supports arrays as one of its data types. It allows for efficient storage of numerical data and metadata, using the same format for both. Unfortunately, the harmonisation of the two happens at the expenses of the overall self-explanatory aspect of the data format, and it requires the overhead of parsing in order to load the data into memory.

Apache \href{https://avro.apache.org}{Avro} is another format that can be compared to a binary version of \texttt{JSON}, but due to its overall complexity and lack of native random access support, it is not optimal for long time series.

Formats such as Alphabet's \href{https://developers.google.com/protocol-buffers}{Protocol Buffers} (and similarly \href{https://google.github.io/flatbuffers/}{FlatBuffers}) are specified by a sophisticated domain-specific language and a dedicated compiler with platform-specific implementations which are required to work with the format. Such a format is not self-explanatory, and extremely difficult to reverse engineer should the required software tools become obsolete.

Hierarchical Data Format (\href{https://www.hdfgroup.org/solutions/hdf5}{HDFS}, currently at version 5) is a widely used format that allows for hierarchical metadata and numerical data, in a single file. It automatically performs run length encoding for more compact representation. The number format can be configured to further optimize the file size. Unfortunately the HDF5 specification is complicated and lengthy, and files are not human-readable. Also, data within a file cannot be removed without reconstructing the entire file \citep{greenfield15}.

Audio formats such as WAV (Microsoft's Wave format)  or Matroska Multimedia Container, can also be used to store digital biosensor data \citep{lambert15}. However, such formats only apply to uniformly-sampled data which essentially rules them out. Their metadata are also inflexible and specific to audio applications. %Furthermore, MPEG audio is generally compressed which introduces error on reconstruction.

\section{Proposed data format} \label{proposal}
Given the above discussion, our proposed format stores metadata and numerical sensor data in separate files. This separation was chosen because different priorities and trade-offs apply to numerical data and metadata, particularly considering storage efficiency, human readability and context description flexibility.

The text metadata format is \texttt{JSON} (UTF-8 encoding), primarily chosen because it is both human and machine readable, and more compact than XML. Figure \ref{recordingMetadata} illustrates the proposed metadata structure, and Table \ref{metadataTable1} and defines each of its fields. Additional example metadata files are given in the Appendix, and corresponding \texttt{JSON} validation schemas are available upon request.

This metadata proposal is focused on a robust definition of the recording itself, i.e the information that is necessary and sufficient for parsing the numerical sensor data into memory, and giving it the correct and unambiguous physical interpretation. This can be then linked to separate metadata tables that define additional context such as subject background, protocol, experiment setup and others. Together, these metadata tables complete all information needed to understand the context of the recording.

\begin{figure*}
\centering
  \includegraphics[width=.9\textwidth]{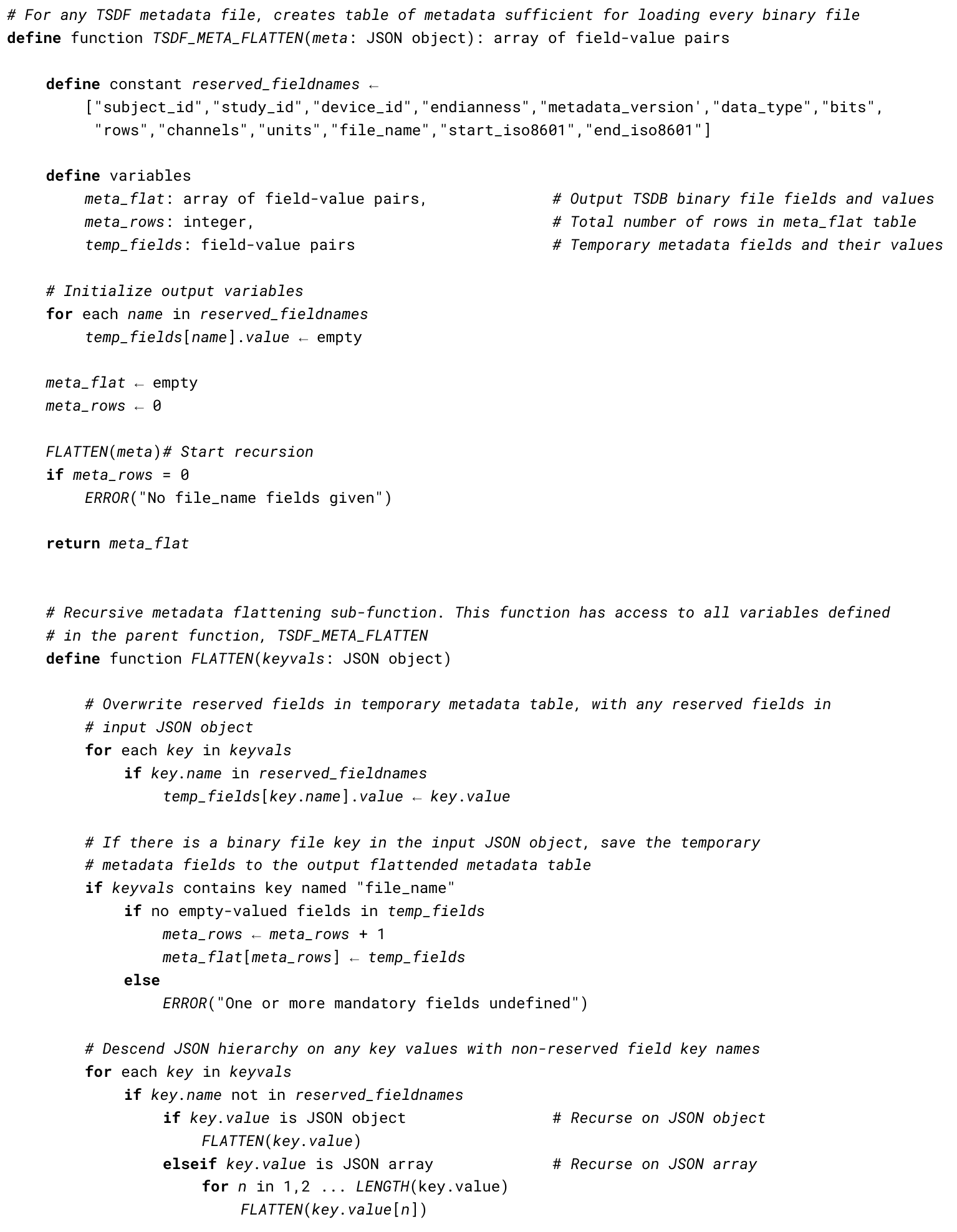}
  \caption{Pseudocode for parsing an arbitrary TSDF format \texttt{JSON} metadata file, in order to retrieve sufficient metadata in order to load every binary file associated with the metadata. The function $TSDF\_META\_FLATTEN$ takes as input a top-level \texttt{JSON} object, and returns an array of field-value pairs, one element in the array per \textbf{file$\_$name} field encountered anywhere in the \texttt{JSON} hierarchy. It recursively calls the FLATTEN function on \texttt{JSON} objects inside other objects and/or arrays of objects.}
  \label{flattenAlgorithm}
\end{figure*}

Fields such as \textbf{study$\_$id}, \textbf{subject$\_$id} and \textbf{device$\_$id} are identifiers that need to be specified in order to correctly link the data to its context. The field \textbf{metadata\_version} ensures correspondence between a metadata file and its JSON validation schema, as this is likely to evolve in future format versions. The fields in bold in table \ref{metadataTable1} are mandatory, but can occur at different levels in the hierarchy. Such flexible \texttt{JSON} schema allows to encode different hierarchical choices in simple flat structures (the example from Figure \ref{recordingMetadata} is reported in Table \ref{flattable}), which work for simple cases with only amplitudes of a single time series without timestamps, but also for multi-subject, multi-modality setting with all timestamps specified (examples in the Appendix in section \ref{app}). One traverses the metadata tree from the root, collecting the relevant fields, until \textbf{file$\_$name} indicates a leaf of the hierarchy has been reached. At this point all mandatory information should be present. If the same field name is encountered multiple times while traversing the hierarchy, the deeper level overwrites the level closer to the root, following the principle of inheritance. When a group of multiple \textbf{file$\_$name}s are found together as a list, this indicates those belong together, for example amplitudes of an accelerometer, amplitudes of a gyroscope and their corresponding timestamps. Figure \ref{flattenAlgorithm} provides pseudocode to implement such flexible metadata structure.

To prevent ambiguities or inconsistencies, redundancy in the metadata is kept to a minimum. The few redundant fields in the specification are included for practical reasons, e.g. both \textbf{start$\_$iso8601} and \textbf{end$\_$iso8601} (e.g. \texttt{"2021-02-20T14:56:37.123+02:00"}) are present in the list of mandatory fields even if the latter can be inferred from the former and the timestamp information. It is, however, convenient to have direct access to the end timestamp of the recording without having to load the entire set of timestamps in order to calculate it. It is recommended to implement consistency checks when storing such redundant information in order to avoid possible mismatches, e.g. checking whether the most recent point in time derived from the binary timestamp information and the start timestamp, indeed correspond to the end timestamp.

For each recording session, represented in one single metadata file, multiple numerical files can be generated, e.g. when different sensor types are involved. For this reason,  multiple \textbf{signals} can be annotated in the metadata (the field consists of an array of instances). Because of non-uniform sampling, often timestamp information needs to be associated with each sample. The data structure of this time information is essentially the same as the amplitude data, both for the numerical data and the metadata. Therefore we use the same metadata structure for both amplitudes and timestamps, grouping time and amplitude information together for each type of sensor under the \textbf{samples} header, specifying \textbf{sensor$\_$type} (e.g. \texttt{"accelerometer"}) for each group. This allows to share timestamp information among several amplitude files (e.g. accelerometer and gyroscope in a single IMU). Each instance of time or amplitude information adheres to the same schema, with fields such as  \textbf{file$\_$name} indicating the corresponding binary sensor data file (e.g. \texttt{"acc$\_$ampl.BIN"}). 

Units are compliant with the International System of Units (SI), more specifically the encoding of HL7's Unified Code for Units of Measure. The field \textbf{units} of the metadata allows a specific unit for each channel in the binary sensor data file, which makes it possible to integrate time values with the measurements (e.g. \texttt{["ms","g","g","g"]}), or to store data from multiple sensor types/derived features in a single binary data file. 
Similarly to the unit of measures, \textbf{channels} need to be specified (e.g. \texttt{["time","X","Y","Z"]}).

For correct loading of the binary data, it is fundamental that the information on \textbf{data\_type} (e.g. \texttt{"float"}) and \textbf{bits} (e.g. \texttt{32}) is present. It is also necessary to indicate which time encoding strategy has been used (e.g. \texttt{"relative"}). This is specified in the \textbf{compression} field of the time section.

For the sensor data itself, raw binary files (\texttt{BIN} format) are used. We explicitly avoid sophisticated compression or serialization formats, because both require complex decoding and/or parsing which severely hampers random access and fast loading of the data into memory, even if they may optimize storage efficiency.

We typically split timestamps and sensor numerical data in separate files, because their data type demands are different. For instance, time information can typically be stored in unsigned, 32 bit integers, assuming millisecond precision, whereas sample measurements may be floating point numbers.

In the case of multiple measurement channels, there are two possibilities for the serialization order of the measurements: either \textit{multiplexed} (e.g.: $x_1$, $y_1$, $x_2$, $y_2\dots$), or \textit{channel-first} (e.g. $x_1$, $x_2 \dots$, $y_1$, $y_2 \dots$). However, multiplexed has the clear advantage of facilitating native random access support, as the values are grouped together by a common timestamp. Therefore, the format always requires multiplexing.

%A new dataset needs to be started under these circumstances:
%\begin{itemize}
%\item for each subject, measurement device and recording session
%\item when changing time zone or recording mode
%\item when a maximum file size is reached: for faster sampling rates this is typically 1 file/day, for slow sampling multiday recordings are possible.
%\end{itemize}

\begin{table*}[t]
\begin{tabular}{ p{.25\linewidth} l p{.65\linewidth} }
  \textbf{Field} & \textbf{Field type} & \textbf{Description}\\
  \hline
  \textbf{subject\_id} & \texttt{string} & Unique ID of the subject, primary key to subject metadata elsewhere\\
 \textbf{study\_id} & \texttt{string} & Unique ID of the study/project/experiment, primary key to study metadata elsewhere\\
  \textbf{device\_id} & \texttt{string} & Unique ID of the measurement device, primary key to device metadata elsewhere\\
  %protocol\_id & \texttt{string} & Unique ID of the protocol description/SOP\\
  %source\_file\_path & \texttt{string} & Path to the original data file, before conversion to this format\\
  \textbf{endianness} & \texttt{string} & Byte order, {\tt "big"} or {\tt "little"} (endian), for numerical values in binary data\\
  \textbf{metadata\_version} & \texttt{string} & Version of the definition of this data format\\
  %checksum\_type & \texttt{string} & Algorithm used to generate checksums/hash code for data integrity, e.g. MD5\\
  \textbf{start\_iso8601} & \texttt{string} & ISO 8601 time stamp for the start of the recording, with ms precision\\ %If case of multimodal recordings, the earliest any of the modalities starts recording\\
  %start\_unix\_ms & \texttt{uint} & Unix time time stamp for the start of the recording, with ms precision, as an alternative to ISO8601\\
  \textbf{end\_iso8601} & \texttt{string} & ISO 8601 time stamp for the end of the recording, with ms precision\\ %In case of multimodal recordings, the latest any of the modalities stops recording\\
  %end\_unix\_ms & \texttt{uint} & Unix time time stamp for the end of the recording, with ms precision, as an alternative to ISO8601\\
  %  signals/sensor$\_$type & \texttt{string} & Name of the sensor modality or the derived features in the file\\
    \textbf{rows} & \texttt{uint} & Total amount of samples per channel, i.e. the number of time points\\
    %signals/approx\_sampling\_freq & \texttt{float} & Convenient to read without calculation from time file (requires consistency check), or needed in case of uniform sampling without timestamps. Expressed in Hz\\
    %signals/samples & \texttt{[]} & for each modality, a set of amplitude and timestamp metadata corresponding to that modality, as specified below\\
    
  \textbf{file\_name} & \texttt{string} & Name of the file containing the amplitude and/or timestamp data\\
  \textbf{channels} & \texttt{string[]} & Labels for each of the channels, e.g. \texttt{["X"}, \texttt{"Y"}, \texttt{"Z"]} for accelerometry data\\
  \textbf{units} & \texttt{string[]} & Unit for each of the channels, e.g.  \texttt{["ms"]} for time data\\
  \textbf{data\_type} & \texttt{string} & Number format of the amplitudes or timestamps, e.g. \texttt{"int"} for integer,  \texttt{float} for floating point\\
  \textbf{bits} & \texttt{uint} & Bit-length of the number format\\
  %\qquad compression & \texttt{string} & Compression method applied to the amplitudes (defaults to none), for time information the time encoding method: \textit{difference, relative} or \textit{absolute} (defaults to \textit{relative})\\
  %\qquad scale\_factors & \texttt{float[]} & Array of the same length as the number of channels containing the factors to apply to convert the amplitudes to the values in the specified units (defaults to 1)\\
  %\qquad checksum & \texttt{string} & Checksum value for the amplitudes or timestamps data\\
\end{tabular}
\caption{\label{metadataTable1} Mandatory fields for metadata files conforming to the proposed format.}
\end{table*}

\begin{table*}[t]
\ttfamily
\begin{tabular}{l|llll}

\textbf{\textrm{Field name}   }                   & \multicolumn{2}{c}{\textbf{\textrm{Field value}}}                                                                             &  &  \\
\hline

{\color[HTML]{4A76F7}file\_name}                               & {\color[HTML]{CB0000}"sensor\_time.bin"}                                      & {\color[HTML]{CB0000}"sensor\_samples.bin"}                                   &  &  \\
{\color[HTML]{4A76F7} subject\_id}       & \cellcolor[HTML]{EFEFEF}"0713"                          & \cellcolor[HTML]{EFEFEF}"0713"                          &  &  \\
{\color[HTML]{4A76F7} study\_id}         & \cellcolor[HTML]{EFEFEF}"drug513trialphase2"            & \cellcolor[HTML]{EFEFEF}"drug513trialphase2"            &  &  \\
{\color[HTML]{4A76F7} device\_id}        & \cellcolor[HTML]{EFEFEF}"serialUID071290123"            & \cellcolor[HTML]{EFEFEF}"serialUID071290123"            &  &  \\
{\color[HTML]{4A76F7} endianess}         & \cellcolor[HTML]{EFEFEF}"little"                        & \cellcolor[HTML]{EFEFEF}"little"                        &  &  \\
{\color[HTML]{4A76F7} metadata\_version} & \cellcolor[HTML]{EFEFEF}"0.1"                           & \cellcolor[HTML]{EFEFEF}"0.1"                           &  &  \\
{\color[HTML]{4A76F7} start\_iso8601}    & \cellcolor[HTML]{EFEFEF}"2019-12-19T12:41:45.716+00:00" & \cellcolor[HTML]{EFEFEF}"2019-12-19T12:41:45.716+00:00" &  &  \\
{\color[HTML]{4A76F7} end\_iso8601}      & \cellcolor[HTML]{EFEFEF}"2019-12-19T13:39:33.151+00:00" & \cellcolor[HTML]{EFEFEF}"2019-12-19T13:39:33.151+00:00" &  &  \\
{\color[HTML]{4A76F7} rows}              & \cellcolor[HTML]{EFEFEF}4833                            & \cellcolor[HTML]{EFEFEF}4833                            &  &  \\
{\color[HTML]{4A76F7} channels}          & {[}"time"{]}                                            & {[}"pos","vel","accl"{]}                                &  &  \\
{\color[HTML]{4A76F7} units}             & {[}"s"{]}                                               & {[}"m","m/s","m/s/s"{]}                                 &  &  \\
compression                              & "difference"                                            &                                                         &  &  \\
{\color[HTML]{4A76F7} data\_type}        & "float"                                                 & "int"                                                   &  &  \\
{\color[HTML]{4A76F7} bits}              & 32                                                      & 16                                                      &  & 

\end{tabular}
\textrm{\caption{\label{flattable}Flattened view of the \texttt{recording\_metadata.JSON} file in Figure \ref{recordingMetadata}. Mandatory fields are shown in blue, and each column of data refers to a different file (names in red). Note that the data in grayed cells are inherited from the root level of the metadata and are therefore common for both files.  }}
\end{table*}

\section{Conclusions}
There is yet no sensor type, species, and disease-independent open standard addressing the specific needs of physiological biosensor data. To address this issue, we propose a simple data storage strategy that consists of structured metadata files in \texttt{JSON} format, and time and measurement data stored in simple, multiplexed, raw binary files. To ensure unambiguous reconstruction of the binary data, the metadata contains a set of mandatory fields that are limited to essential sensor measurement information. These metadata files are both human and machine readable in order to facilitate querying and high efficiency, random access loading of the sensor measurement data. Storage and retrieval of other, relevant metadata (such as device, subject and protocol information) is made possible through unique identifiers acting as key indices into associated relational tables. An implementation of a software tool to create and load data in this novel format, accompanies this paper.%Further effort is needed for defining the common features of such files and incorporating them into this proposed format without losing generality.

\section{Acknowledgments}
We are grateful to the Michael J Fox Foundation for Parkinson’s Research and UCB Biopharma.

\clearpage

\begin{table*}
\section{Appendix}
\label{app}

An example of a flat metadata structure for a single audio recording.\\

\begin{verbatimtab}
{
   "study_id": "voicedata",
   "subject_id": "recruit089",
   "device_id": "audiotechnica02",
   "endianness": "little",
   "metadata_version": "0.1",
   "start_iso8601": "2016-08-09T10:31:00.000+00:00",
   "end_iso8601": "2016-08-10T10:31:30.000+00:00",
   "sampling_rate": 44100,
   "rows": 1323000,
   "channels": ["left","right"],
   "units": ["unitless","unitless"],
   "compression": "none",
   "data_type": "int",
   "bits": 16,
   "file_name": "audio_voice_089.raw"
}


\end{verbatimtab}
\end{table*}

\begin{table*}

An example of a two-level hierarchical metadata structure, embedding two recordings from a device with accelerometry and temperature sensors.\\
\begin{verbatimtab}
{
   "subject_id": "PD0234",
   "study_id": "homestudy22",
   "device_id": "XBT7456",
   "endianness": "little",
   "metadata_version": "0.1",
   "data_type": "float",
   "bits": 32,
   "multi-day_session": [
      {
         "start_iso8601": "2022-26-10T09:26:45.123+00:00",
         "end_iso8601": "2022-26-10T09:36:52.266+00:00",
         "bits": 32,
         "sensors": [
            {
               "rows": 60714,
               "file_name": "accelerometer_t1.bin",
               "channels": ["time,","magnitude"],
               "units": ["ms","m/s/s"]
            },
            {
               "rows": 607,
               "file_name": "temperature_t1.bin",
               "channels": ["time","temperature"],
               "units": ["s","deg_C"]
            }
         ]
      },
      {
         "start_iso8601": "2022-28-10T10:42:12.465+00:00",
         "end_iso8601": "2019-28-10T13:54:36.578+00:00",
         "sensors": [
            {
               "rows": 1154411,
               "file_name": "accelerometer_t2.bin",
               "channels": ["time","magnitude"],
               "units": ["ms","m/s/s"]
            },
            {
               "rows": 11544,
               "file_name": "temperature_t2.bin",
               "channels": ["time","temperature"],
               "units": ["s","deg_C"]
            }
         ]
      }
   ]
}
\end{verbatimtab}
\end{table*}
\begin{comment}
    "subject_id": "b8f4312212be6b5b",
    "device_id": "7A12b90123",
    "study_id": "bc9d936030c909",
    "protocol_id": "PD@home", 
   
    "source_file_path": "/some/path/to/rawfilename.xyz",
    "endianness": "little",
    "metadata_version": "0.1",
    "checksum_type": "md5",
    
    "start_iso8601": "2019-12-19T12:41:45.716+00:00",
    "end_iso8601":   "2019-12-19T13:39:33.151+00:00",
	
    "signals": [{
		"sensor_type":"accelerometer",
		"rows":483133,
		"approx_sampling_freq":100,
		"samples":[{
			"file_name":"WatchData.IMU.Week0.raw_segment0006_samples.bin",
			"channels": ["X","Y","Z"],
			"units": ["m/s/s","m/s/s","m/s/s"],	
			"data_type": "float",
			"bits": 32,
			"compression": "none", 
			"scale_factors": [0.00469378,0.00469378,0.00469378], 
			"checksum": "8075f7235b0144f34653b149d6065835", 
			},
			{
			"filename": "WatchData.IMU.Week0.raw_segment0006_time.bin",
			"units":["ms"],
			"data_type":"uint",				
			"bits":32,
			"compression":"difference",
			"checksum":"9075f7235b0144f34653b149d6065835"
			}]
		}
	]
\end{comment}

\clearpage
\bibliographystyle{plainnat}
\bibliography{paper}
\end{document}